\documentclass[aps,preprint]{revtex4}%
\usepackage{amsfonts}
\usepackage{amsmath}
\usepackage{amssymb}
\usepackage{graphicx}%
\begin{document}
\title{{\LARGE ON \ THE \ MAGNETIC FIELD, AND ENTROPY \ INCREASE, \ IN \ A \ MACHIAN
UNIVERSE}}
\author{Marcelo Samuel Berman$^{1}$}
\affiliation{$^{1}$Editora Albert Einstein \ Ltda\ - Av. Candido Hartmann, 575 - \ \# 17}
\affiliation{80730-440 - Curitiba - PR - Brazil}

\pacs{04.20.-q ; 04.20.Jb \ ; \ 98.80.-k \ \ ; \ \ 98.80.Jk}
\date{( v.2 ) 19 November, 2006}
\keywords{Cosmology; Einstein; Brans-Dicke; Cosmological term; Planck's Universe; Mach;
Magnetic Field; Entropy; Radiation.}
\begin{abstract}
By means of the experimental result on the present equipartition between
background microwave radiation energy and that of the interstellar magnetic
field, and by advancing a Machian relation for the magnetic field, which,
differently than in other authors' papers, is valid for the entire spanlife of
the Universe, implying that the magnetic field depends on the inverse radius
of the Universe, we obtain a general formula such that \ $B$\ \ depends on
\ $R^{-1}.$Our \ estimate for Planck's magnetic field is 10$^{-3}$ times the
Sabbata and Sivaram's one. It is shown that the energy densities involved in
the above problem are dependent on $R^{-2}$ . For radiation, this implies, as
we show, that the total entropy of the Universe is increasing with expansion.
In particular we show that $\ R\propto T^{-2}$\ , where \ $T$\ \ is the
absolute temperature, and the entropy is proportional to \ $R^{\frac{3}{2}}$\ .\ 

\textbf{Keywords}: Cosmology; Einstein; Brans-Dicke; Cosmological term;
Planck's Universe; Mach; Magnetic Field; Entropy; Radiation.

\textbf{PACS}: \ 04.20.-q ; 04.20.Jb \ ; \ 98.80.-k \ \ ; \ \ 98.80.Jk

\end{abstract}
\maketitle

\begin{center}
{\LARGE ON \ THE \ MAGNETIC FIELD, AND ENTROPY \ INCREASE, \ IN \ A \ MACHIAN
UNIVERSE}

\bigskip

Marcelo Samuel Berman
\end{center}

\bigskip

\bigskip In two recent chapters of edited books (Berman, 2006; 2006a) it was
proposed a new interpretation for Brans-Dicke relation, which, instead of
being an aproximate relation only valid for the present Universe, should be
interpreted as meaning that the mass \ $M$\ \ ,\ of the causally related
Universe, is directly proportional to the radius \ $R$\ \ , in the entire life
of the Universe. In the same references, it is shown that, the new
interpretation of Brans-Dicke relation, along with the hypothesis that the
cosmological "constant" varies with \ $R^{-2}$\ ,\ \ arise from the imposition
that the total energy of the Universe, is zero-valued. Sabbata and
Sivaram(1994), have shown that, in analogy with Brans-Dicke aproximate
relation, one could state another similar one for the spin of the Universe
\ $L$\ \ . Again, Berman(2006b), extending his conjectures on the zero-total
energy of the Universe, and including in the total energy, a term representing
the rotational energy, derived Sabbata and Sivaram(1994) aproximate relation,
as an exact formula, indicating that \ $L$\ \ varied with \ $R^{2}$\ during
all times. In all cases, Berman has made the hypothesis, that the fraction of
each kind of energy participation, did not vary with time, when taken as
fractions of \ $Mc^{2}$\ .\ This implies, \ that if the Universe is
\ $\Lambda$\ -- driven today, it was also along all periods of time. This
"explains"\ \ that the result obtained by \ A. Riess and colaborators,
described by the international media, as of November, 2006, only means that
our Universe is indeed, a zero -- total -- energy entity.

\bigskip

We now extend Berman's hypotheses, while keeping a new term which contributes
to the total energy of the Universe, dictated by the magnetic field. The
fraction of magnetic energy participation to the total energy, is nevertheless
kept in a \ $10^{-3}$\ orders of magnitude, because we take for granted that
the observed equipartition between the microwave background radiation, and
magnetic field energies, for interstellar media, point out to a similar
fraction in the magnetic field of the Universe. We impose that such fraction
endures for the entire history of the Universe; in fact, this means that we
adjust our Machian relation for the magnetic field, in order that its present
value\ should be around \ $10^{-6}$\ Gauss.\ 

\bigskip

As the fractions of energy, of any kind, in the Machian Universe, according to
our theory, are to be maintained, we take for granted, that any kind of
energy's density, varies with \ $R^{-2}$\ , as has been shown, for the total
energy density, by Berman(2006; 2006a) and Berman and Marinho Jr(2001).\ For
each type of energy, we would have a constant fraction of the total energy,
i.e., constant in time.\ For the total energy density, we would have: \ \ 

\bigskip

$\rho=\frac{M}{\frac{4}{3}\pi R^{3}}$\ \ \ \ \ \ \ \ . \ \ \ \ \ \ \ \ \ \ \ \ \ \ \ \ \ \ \ \ \ \ \ \ \ \ \ \ \ \ \ \ \ \ \ \ \ \ \ \ \ \ \ \ \ \ \ \ \ \ \ \ \ \ \ \ \ \ \ \ \ \ \ \ \ \ \ \ \ \ \ \ (1)

\bigskip

From Brans-Dicke relation, as modified by Berman, we have:

\bigskip

$\frac{GM}{Rc^{2}}=\gamma=$ \ \ constant \ $\sim1$\ \ \ . \ \ \ \ \ \ \ \ \ \ \ \ \ \ \ \ \ \ \ \ \ \ \ \ \ \ \ \ \ \ \ \ \ \ \ \ \ \ \ \ \ \ \ \ \ \ \ \ \ \ \ \ (2)

\bigskip

From (1) and \ (2), we obtain the desired dependence, \ \ $\rho\propto R^{-2}$\ \ .

\bigskip

The energy density associated with a magnetic field \ $B$\ \ is given by:

\bigskip

$\rho_{B}=\frac{B^{2}}{8\pi}$\ \ \ \ \ \ \ \ \ \ \ \ \ . \ \ \ \ \ \ \ \ \ \ \ \ \ \ \ \ \ \ \ \ \ \ \ \ \ \ \ \ \ \ \ \ \ \ \ \ \ \ \ \ \ \ \ \ \ \ \ \ \ \ \ \ \ \ \ \ \ \ \ \ \ \ \ \ \ \ \ \ (3)

\bigskip

In mass units, we have to divide the second member of (3) by \ $c^{2}$\ . The
total energy fraction for the magnetic field, relative to \ $Mc^{2}$\ would be
given by:

\bigskip

$\left[  \frac{4}{3}\pi R^{3}\right]  \left[  \frac{B^{2}}{8\pi c^{2}}\right]
\left[  Mc^{2}\right]  ^{-1}=\gamma_{B}\cong10^{-6}$\ \ \ \ \ \ \ \ . \ \ \ \ \ \ \ \ \ \ \ \ \ \ \ \ \ \ \ \ \ \ \ \ \ \ \ \ \ \ \ (4)

\bigskip

We then find that \ $B\propto R^{-1}$\ \ because, in fact, from (4) we have:

\bigskip

$B^{2}=12$ $\ c^{4}$ $\gamma$ $\gamma_{B}$ $G^{-1}R^{-2}$\ \ \ \ \ \ \ \ \ . \ \ \ \ \ \ \ \ \ \ \ \ \ \ \ \ \ \ \ \ \ \ \ \ \ \ \ \ \ \ \ \ \ \ \ \ \ \ \ \ \ \ \ \ \ \ \ \ \ \ (5)

\bigskip

We then find, for the present Universe, with \ $R\cong10^{28}$\ cm,
\ $B\cong10^{-6}$\ Gauss\ .

\bigskip

For Planck's Universe, we would have, with \ $R_{Pl}\cong10^{-33}$\ cm, \ \ 

$\bigskip$

$B_{Pl}=B\left[  \frac{R}{R_{Pl}}\right]  \cong10^{55}$\ Gauss.\ \ 

\bigskip

This last value is larger than the maximum limit for the magnetic field not to
provoke instabilities in the vacuum, according\ \ to a recent analysis made
through Quantum Electrodynamics theory (QED), by Shabad and Usov(2006). That
being the case, we can imagine this fact as causing the eruption of the
inflationary phase, immediatly after Planck's time.

\bigskip\bigskip We remark that Sabbata and Sivaram(1994) obtained, in other
context, for the Planck's magnetic field, \ the value\ \ $10^{58}$\ Gauss,
which is larger than in our estimate.

\bigskip

Though we have derived the dependency of the magnetic field with \ $R^{-1}$\ ,
from the zero-total energy conjecture, in a Machian Universe, (see relation 4
above), this is also obtained from the energy density dependence on \ $R^{-2}%
$\ , which is also derived from the above conjecture, as also has been
considered above, for any kind of particular energy, during the lifespan of
the Universe.\ \ 

\bigskip

It has been argued by an annonymous referee, that there would be one kind of
energy density, that would not obey the \ $R^{-2}$\ \ - dependence, \ namely,
the radiation one. Though the radiation energy density obeys the well-known
black body law \ 

$\bigskip$

$\rho_{R}=aT^{4}$\ \ \ \ \ \ \ \ \ \ ( \ $a$ \ = constant\ )
\ \ \ \ \ \ \ \ \ \ , \ \ \ \ \ \ \ \ \ \ \ \ \ \ \ \ \ \ \ \ \ \ \ \ \ \ \ \ \ (6)

\bigskip

where \ $T$\ \ stands for the absolute temperature, we must first agree on the
dependence between \ $R$\ \ \ and \ \ $T$\ \ \ . In standard Cosmology, it is
accepted as a general rule, that the total entropy of the Universe is constant
in time; then, it is easily shown that

\bigskip

\ $RT=$\ \ constant. \ \ \ \ \ \ \ \ \ \ \ \ \ \ \ \ \ \ \ \ \ \ \ \ \ \ \ \ \ \ \ \ \ \ \ \ \ \ \ \ \ \ \ \ \ \ \ \ \ \ \ \ \ \ \ \ \ \ \ \ \ \ \ \ \ \ \ \ (7)

\bigskip

Notwithstanding, this hypothesis for the entropy has been troubled by several
observations, which run from the verification that photons' production in the
Universe, must be accompanied\ \ by an increasing value for the entropy of the
Universe. Weinberg (1972), has long ago warned that the perfect fluid
hypothesis concerning the non-dissipative fluid, is problematic. He even
advanced the idea of a viscous fluid. It is well-known that a perfect fluid is isentropic.

\bigskip

If the Machian perspective is taken into account, we should have:

\bigskip

$\rho_{R}=\beta R^{-2}$ \ \ \ \ \ \ \ \ ( \ $\beta$\ \ = constant )\ \ \ \ \ . \ \ \ \ \ \ \ \ \ \ \ \ \ \ \ \ \ \ \ \ \ \ \ \ \ \ \ \ \ \ \ \ \ \ \ \ (8)

\bigskip

A comparison between (6) and (8), makes us believe that \ $\ R\propto T^{-2}%
$\ \ ; this relation is obtainable from the zero-total energy conjecture for
the Universe. If we write the energy as represented by an inertial term
($Mc^{2}$) minus the potential energy ($\frac{GM^{2}}{2R}$) plus the radiation
energy \ ($\frac{4}{3}\pi R^{3}\rho_{R}$)\ \ , and equate the result to zero,
we find:

\bigskip

$\frac{GM}{c^{2}R}-\frac{4\pi R^{3}}{3Mc^{2}}\rho_{R}\cong1$
\ \ \ \ \ \ \ \ \ \ \ \ . \ \ \ \ \ \ \ \ \ \ \ \ \ \ \ \ \ \ \ \ \ \ \ \ \ \ \ \ \ \ \ \ \ \ \ \ \ \ \ \ \ \ \ \ \ \ \ \ \ \ \ \ (9)

\bigskip

From Brans-Dicke original relation(Brans and Dicke, 1961), we find relation
(2). This suggests that we also should have the following relation:

\bigskip

$\frac{4\pi R^{3}\rho_{R}}{3Mc^{2}}=\gamma_{R}=$ constant $\sim1$
\ \ \ \ \ \ \ \ \ \ \ \ \ . \ \ \ \ \ \ \ \ \ \ \ \ \ \ \ \ \ \ \ \ \ \ \ \ \ \ \ \ \ \ \ \ \ \ \ \ \ \ \ \ (10)

\bigskip

From the above, we obtain:

\bigskip

$\frac{aGT^{4}R^{2}}{c^{4}}=$ \ constant \ \ \ \ \ \ \ \ \ \ \ \ . \ \ \ \ \ \ \ \ \ \ \ \ \ \ \ \ \ \ \ \ \ \ \ \ \ \ \ \ \ \ \ \ \ \ \ \ \ \ \ \ \ \ \ \ \ \ \ \ \ \ \ \ \ \ \ (11)

\bigskip

We conclude that \ $\ R\propto T^{-2}$\ \ as \ we believed above. This
dependence was studied earlier for non-relativistic decoupled matter (Kolb and
Turner, 1990).

\bigskip

We now check the total entropy of this (Machian) Universe:

\bigskip

$S\propto sR^{3}\propto\frac{\rho_{R}}{T}R^{3}\propto T^{3}R^{3}\propto
R^{\frac{3}{2}}$ \ \ \ . \ \ \ \ \ \ \ \ \ \ \ \ \ \ \ \ \ \ \ \ \ \ \ \ \ \ \ \ \ \ \ \ \ \ \ \ \ \ (12)

\bigskip

In the above, $S$ and \ $s$ \ stand respectivelly for the total entropy, and
the entropy density; from thermodynamics, $S=\frac{E_{R}}{T}$ \ , where
\ \ $E_{R}$\ \ stands for the total radiational energy.

\bigskip

From (12), we have the result that should had been obvious to cosmologists
long ago: total entropy increases with the expanding Universe.

\bigskip

The above result is approximately similar to the one of black hole
thermodynamics: the entropy, in this case, is proportional to the surface area
of the event horizon, \ \ $4\pi R^{2}$\ \ . Here \ \ $R$\ \ stands for
Schwarzschild's radius, $2GM$ \ .\ \ Notice that in the expanding Universe,
the event horizon radius also should grow with time, due to the isotropy and
homogeneity of the Universe, which causes increasing entropy for black holes,
too. This is a Classical Physics effect, independent of Quantum theory.

\bigskip

The perfect fluid hypothesis, for the Universe, is here challenged by Mach's
Principle. We conclude that, because we did not restrict ourselves, to a
specific gravitational theory, like General Relativity, Brans-Dicke or other
alternative one theories, any theory can be Machian, provided that the above
framework is accomodated in those theories.

\bigskip

{\Large Acknowledgements}

\bigskip

The author gratefuly thanks his intellectual mentors, Fernando de Mello Gomide
and M. M. Som, and is also grateful for the encouragement by Albert, Paula and
Geni, and by Marcelo Guimar\~{a}es, Nelson Suga, Mauro Tonasse, Antonio
Teixeira and Herman J. M. Cuesta. Comments by Dimi Chalakov, on several papers
of mine, are highly praised.

\bigskip

\bigskip

{\Large References}

\bigskip

\bigskip Berman,M.S. (2006) - \textit{Energy of Black Holes and Hawking's
Universe -} in \ \textit{Trends in Black Hole Research,} ed. by Paul Kreitler,
Nova Science, New York.

Berman,M.S. (2006 a) - \textit{Energy, Brief History of Black Holes, and
Hawking's Universe}, in \textit{New Developments in Black Hole Research,} ed.
by Paul Kreitler, Nova Science, New York.

\bigskip Berman,M.S. (2006 b) - \textit{On the Machian Properties of the
Universe} - submitted. (Also posted in Los Alamos archives, http://arxiv.org/abs/physics/0610003).

\bigskip Brans, C.; Dicke, R.H. (1961) - Physical Review, \textbf{124}, 925.

Kolb, E.W.; Turner, M.S. (1990) - \textit{The Early Universe}, Addison-Wesley, N.Y.

Sabbata, V.de; Sivaram, C. (1994) - \textit{Spin and Torsion in Gravitation} -
World Scientific, Singapore.

Shabad, A.E.; Usov, V.V. (2006) - Phys. Rev. Lett. \textbf{96}, 180401.

\bigskip Weinberg, S. (1972) - \textit{Gravitation and Cosmology}, Wiley, New York.

\end{document}